\def\hetetwo{HETE-2\ }
\def\hetetwonosp{HETE-2}
\def\swift{{\it Swift\ }}

\def\bsax{{\it Beppo}SAX\ }

\documentclass[
    ,final            
  ]
  {aipproc}

\layoutstyle{6x9}

\begin{document}

\title{GRB Redshift Evolution Within the Unified Jet Model}

\author{T. Q. Donaghy}{
  address={Department of Astronomy \& Astrophysics, University of
Chicago, Chicago, IL 60637}
  ,altaddress={quinn@oddjob.uchicago.edu}
}

\author{D. Q. Lamb}{
  address={Department of Astronomy \& Astrophysics, University of
Chicago, Chicago, IL 60637}
}

\author{C. Graziani}{
  address={Department of Astronomy \& Astrophysics, University of
Chicago, Chicago, IL 60637}
}

\begin{abstract}

\hetetwo has provided new evidence that gamma-ray bursts may evolve
with redshift \citep{graziani2003}.  We investigate the consequences of
this possibility for the unified jet model of XRFs and GRBs
\citep{lamb2003}.  We find that burst evolution with redshift can be
naturally explained within the unified jet model, and the resulting
model provides excellent agreement with existing \hetetwo and \bsax
data sets.  In addition, this evolution model produces reasonable fits
to the BATSE peak photon number flux distribution -- something that
cannot be easily done without redshift evolution.

\end{abstract}

\maketitle

\section{Introduction}

Most objects at cosmological distances (stars, galaxies, AGN) display
evolution of their observable properties with redshift.  Since
gamma-ray bursts (GRBs) are thought to originate in the core-collapse
of massive stars and are observed over a wide range in redshift, it is
reasonable to suppose that they might also evolve.  

Since the advent of rapid GRB localizations with \bsax and
\hetetwonosp, and the consequent follow-up observations, over $30$ GRBs
have reported redshift measurements.  Using 9 \bsax bursts from
\cite{amati2002}, plus an additional 11 bursts localized with
\hetetwonosp, \cite{graziani2003} have been able to strengthen earlier
indications that GRBs are brighter at higher redshifts.

A uniform-jet model proposed by \cite{lamb2003} has been shown to
provide a unified picture of GRBs and XRFs.  Considering all bursts to
have a ``standard energy'' but a range of jet opening solid-angles that
spans five orders of magnitude, this unified jet model can account for
the full range of observed burst properties seen by \hetetwo.  Here we
extend this unified jet model to account for redshift evolution and
show that it can also explain the observed properties of BATSE bursts.

\section{Observations of GRB Evolution}

Analysis of the BATSE catalog has revealed evidence that GRBs may
evolve strongly with redshift.  The use of redshift estimators based on
burst variability has provided evidence that bursts are intrinsically
brighter at larger redshifts than at smaller \cite{lloyd2002,
reichart2001b}.  Analyzing a set of 9 GRBs with spectroscopically
determined redshifts observed by the BeppoSAX satellite,
\cite{amati2002} also claim evidence for an increase in the
isotropic-equivalent energy ($E_{\rm iso}$) with redshift, but they did
not include a discussion of possible threshold selection effects.

Recent results from HETE-2 strengthen the evidence for this
relationship.  Figure 2 from \cite{graziani2003} shows the
isotropic-equivalent energies $E_{\rm iso}$ and luminosities $L_{\rm
iso}$ as a function of redshift for the HETE-2 and BeppoSAX events. 
After correcting for threshold effects, \cite{graziani2003} find that
$E_{\rm iso}$ is correlated with redshift at the $5.1$\% confidence
level, and $L_{\rm iso}$ is correlated with redshift at the $0.9$\%
confidence level.  The observed relationship goes roughly as $E_{\rm
iso} \sim (1+z)^{3}$.

\section{Unified Jet Model Simulations}

The unified jet model of GRBs and XRFs \cite{lamb2003} provides a
natural explanation for redshift evolution in GRBs.  Namely, each burst
exhibits a ``standard energy'' \cite{frail2001,bloom2003}, but the
possible range of jet opening angles varies from fairly large values at
low redshift, to very small values at high redshift, according to the
relationship given above.  That is, evolution in $E_{\rm iso}$ is
explained by an evolution of the jet opening solid-angle, $\Omega _{\rm
jet}$.

\begin{figure}[t]
\includegraphics[height=.3\textheight]{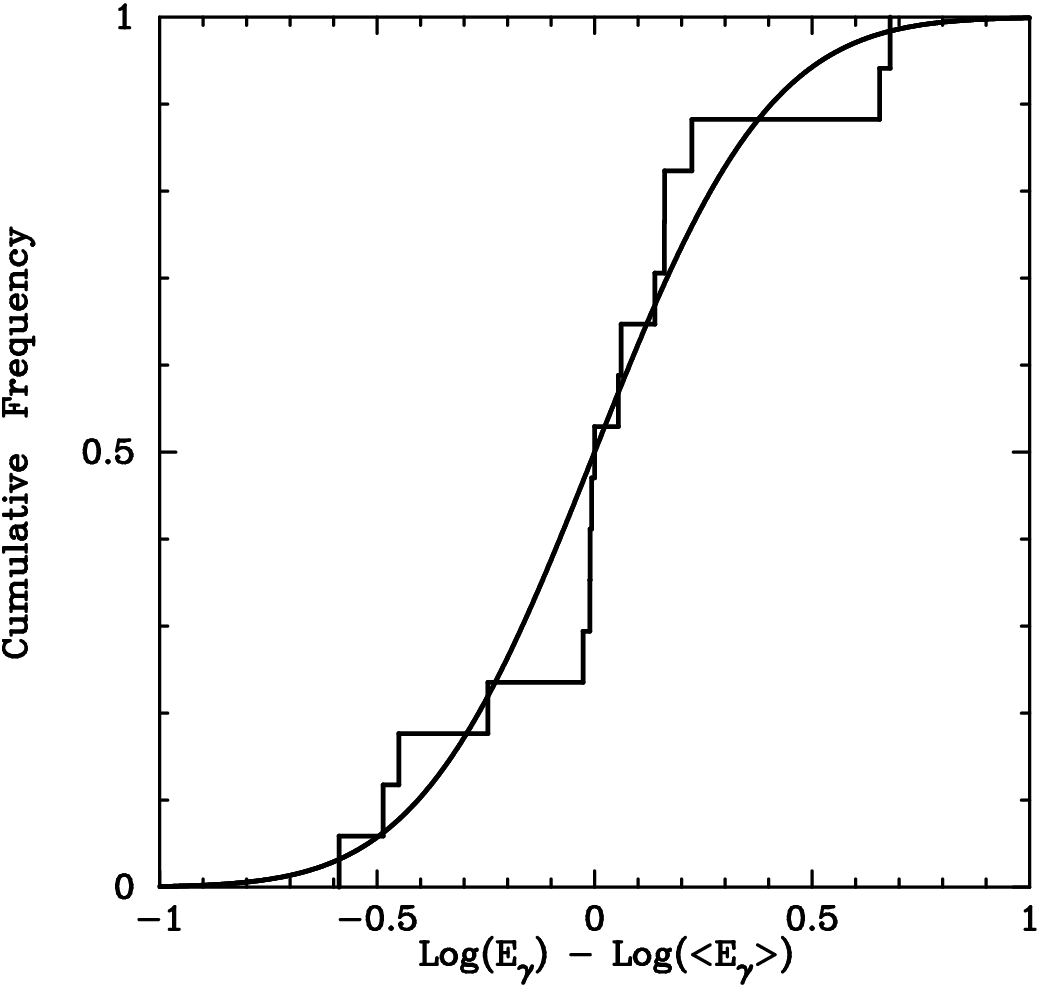}
\includegraphics[height=.3\textheight]{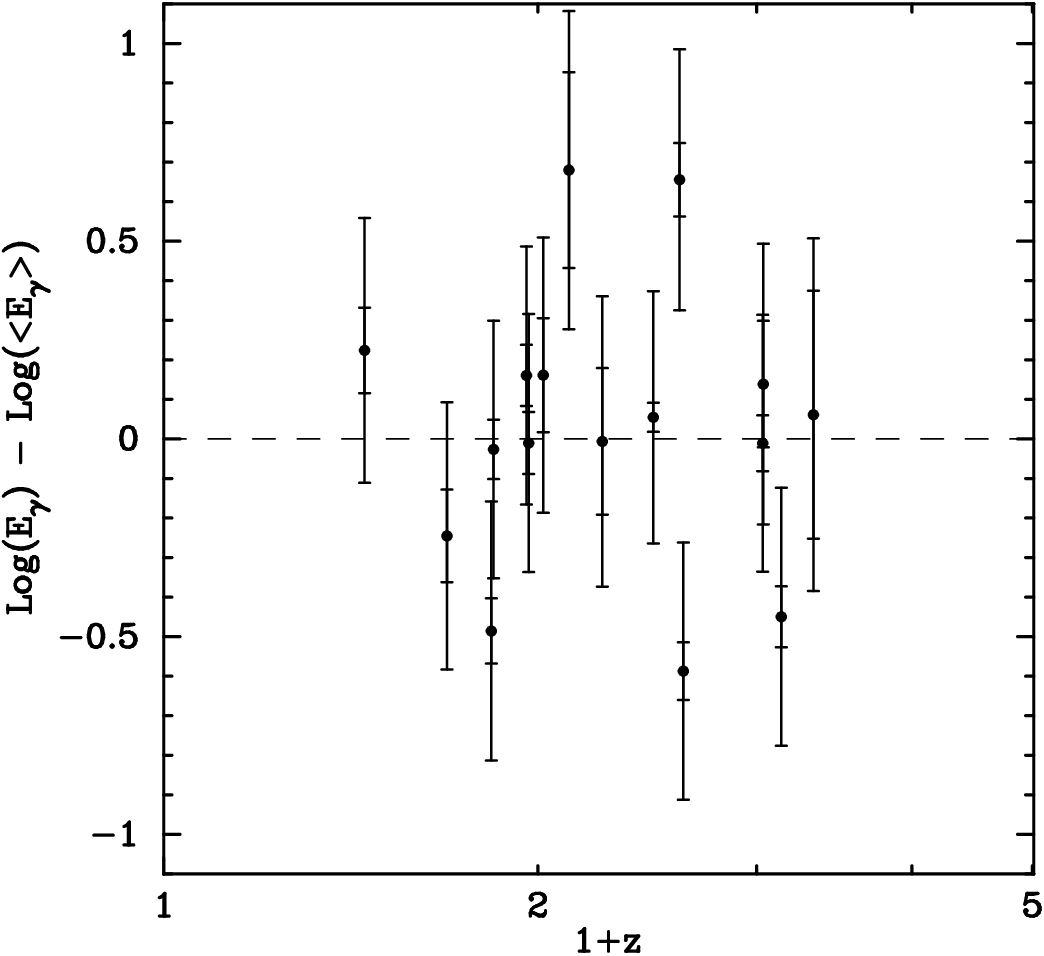}
\caption{{\it Left:} Cumulative distribution of $E_{\gamma}$ values
reported by \cite{bloom2003}, shown with the best-fit Gaussian.  {\it
Right:} Displacement of $E_{\gamma}$ values from the best-fit central
value as a function of redshift.  From \cite{lamb2003}.}
\label{egamma}
\end{figure}

\begin{figure}[t]
\rotatebox{270}{\includegraphics[height=.3\textheight]{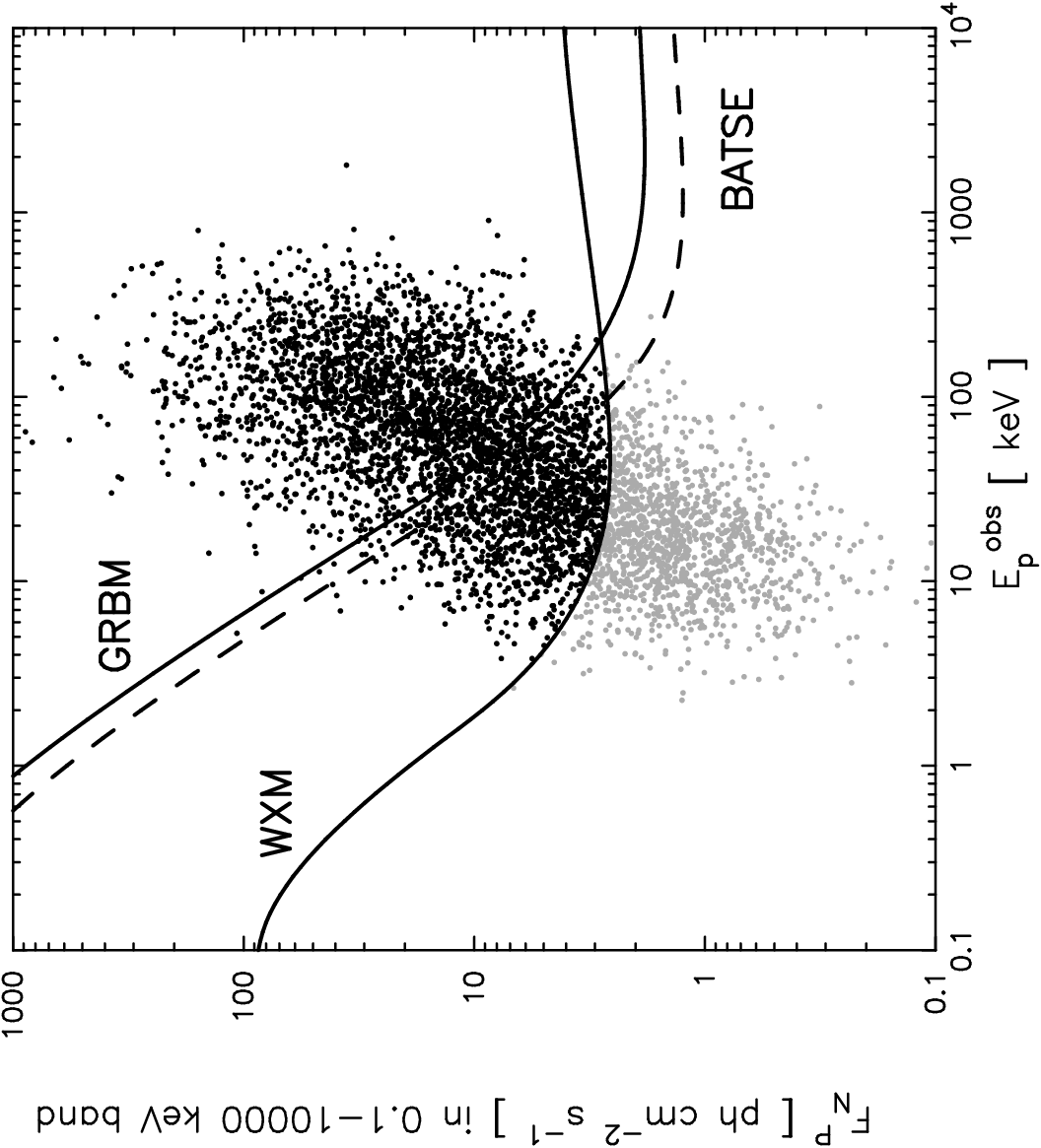}}
\rotatebox{270}{\includegraphics[height=.3\textheight]{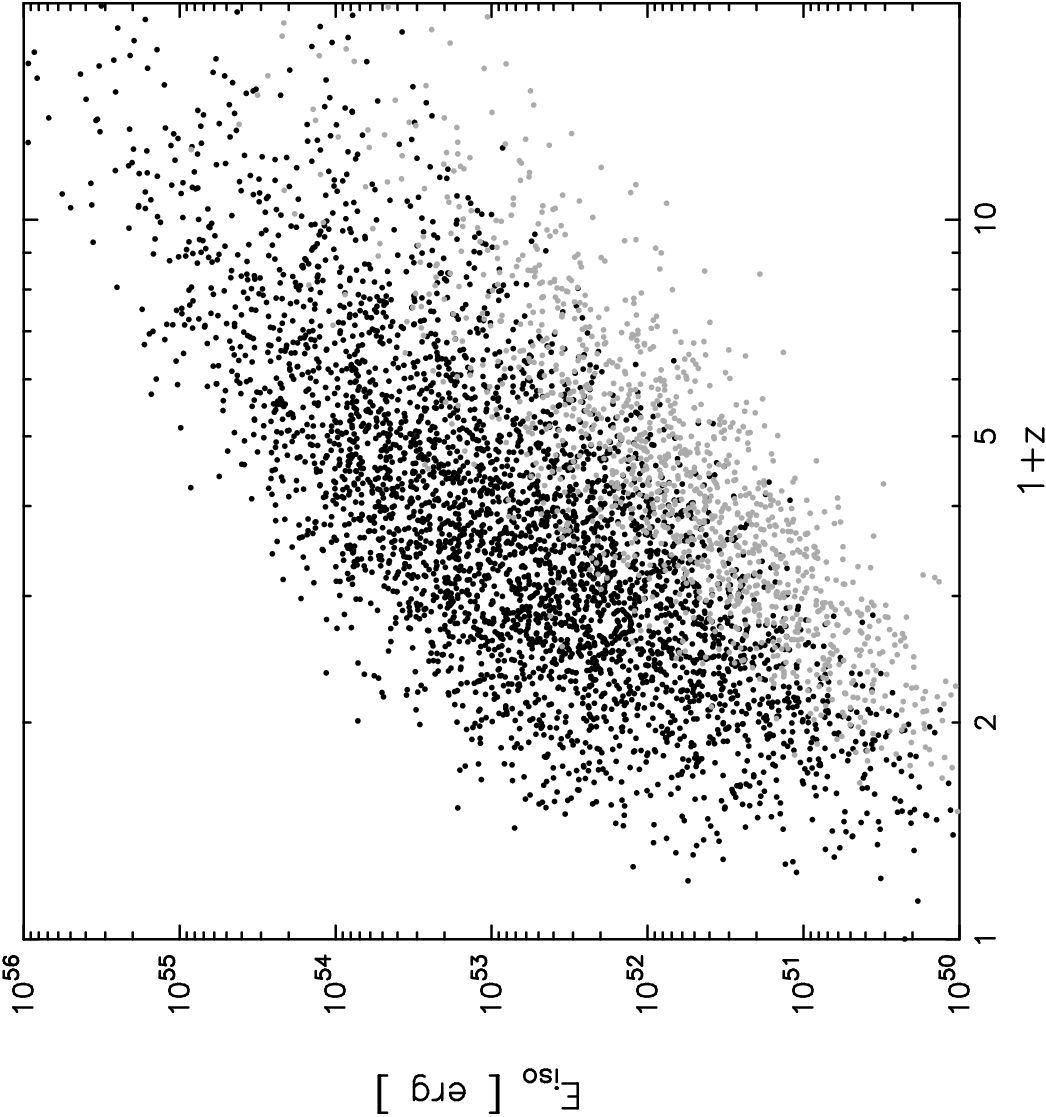}}
\caption{{\it Left:} Distribution of bursts in the ($E_{\rm peak}^{\rm
obs}$, $F_{\rm N}^{\rm P}$)-plane, showing the threshold curves we use
to determine if a burst is detected by various instruments.  Black
points are bursts detected by the WXM, while gray points are not
detected.  {\it Right:} Distribution of bursts in the ($1+z$, $E_{\rm
iso}$)-plane for our best uniform jet with redshift evolution model.}
\label{z1_eiso}
\end{figure}

The burst simulations that we have implmented to test the unified jet
model also provide a powerful way to explore models including burst
evolution with redshift.  For each burst, we obtain: (1) A redshift z
by drawing from a model of the star-formation rate \cite{rr2001}, and
(2) a jet-opening solid angle $\Omega _{\rm jet}$ by drawing from
specific distribution range in $\Omega _{\rm jet}$ that is fixed at
$z=0$ and shifts to smaller values at higher redshifts.  We also
introduce three Gaussian smearing functions to generate: (1) A spread
in jet energy ($E_{\gamma}$, see Figure \ref{egamma}), (2) a spread in
$E_{\rm peak}$ around the $E_{\rm iso}$-$E_{\rm peak}$ relation, and
(3) a spread in the timescale T that converts fluence to flux. Using
these five quantities, we calculate various rest-frame quantities
($E_{\rm iso}$, $E_{\rm peak}$, etc.), and finally, we construct a Band
function for each burst and transform it to the observer frame, which
allows us to calculate fluences and peak fluxes and determine if the
burst would be detected by various instruments (see Figure
\ref{z1_eiso}a).

To obtain the model presented here, we sought to roughly match the
observed distribution in the ($1+z$, $E_{\rm iso}$)-plane (compare
Figure \ref{z1_eiso}b with Figure 2 of \cite{graziani2003}), which
displays a range in $E_{\rm iso}$ of $3\times 10^{49}$ to $1.5\times
10^{52}$ ergs at $z=0$ and $6.4\times 10^{51}$ to $3.2\times 10^{54}$
ergs at $z=5$.  Assuming the faintest burst at $z=0$ corresponds to
$\Omega_{\rm jet}=2\pi$, this translates into a range of $\Omega_{\rm
jet}$ of $2\pi$ to $0.0125$ steradians at $z=0$ and $0.0291$ to
$5.79\times 10^{-5}$ steradians at $z=5$.

The observed values of $E_{\gamma}$ are taken from \cite{bloom2003},
and their distribution is well-fit by narrow Gaussian \cite{frail2001,
bloom2003, lamb2003b} (see Figure \ref{egamma}a).  Figure \ref{egamma}b
plots the displacement of these values from the central value as a
function of redshift and shows no evidence for evolution of
$E_{\gamma}$ with redshift \cite{lamb2003}.  This rules out the
possibility that redshift evolution might be explained by an evolution
of the ``standard energy''.

Since it relies on the random distribution of burst jet axes with
respect to the viewing angle, the universal structured-jet model cannot
easily accommodate redshift evolution.  The only solution would be for
the ``jet energy'', $E_{\gamma}$, to evolve with redshift, but that
would make it difficult to explain Figure 1 and the results of
\cite{frail2001, bloom2003}.

\section{Results}

Figure \ref{cuml_dist} compares the cumulative distributions of four
observable burst quantities against several possible models.  We find
that the uniform jet model with burst evolution can adequately describe
the observed distributions of localized bursts.

\begin{figure}[htb]
\rotatebox{270}{\includegraphics[height=.3\textheight]{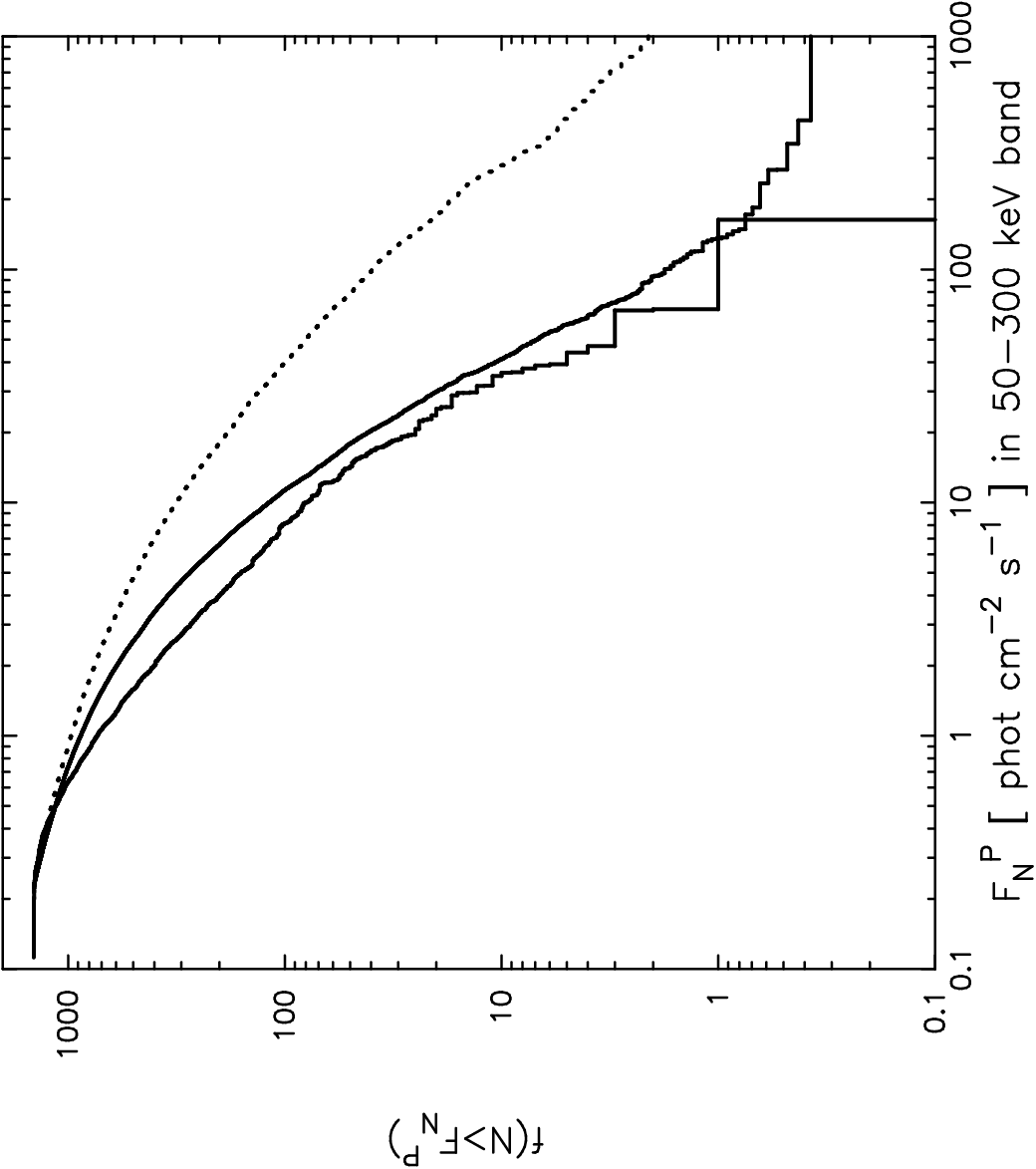}}
\rotatebox{270}{\includegraphics[height=.3\textheight]{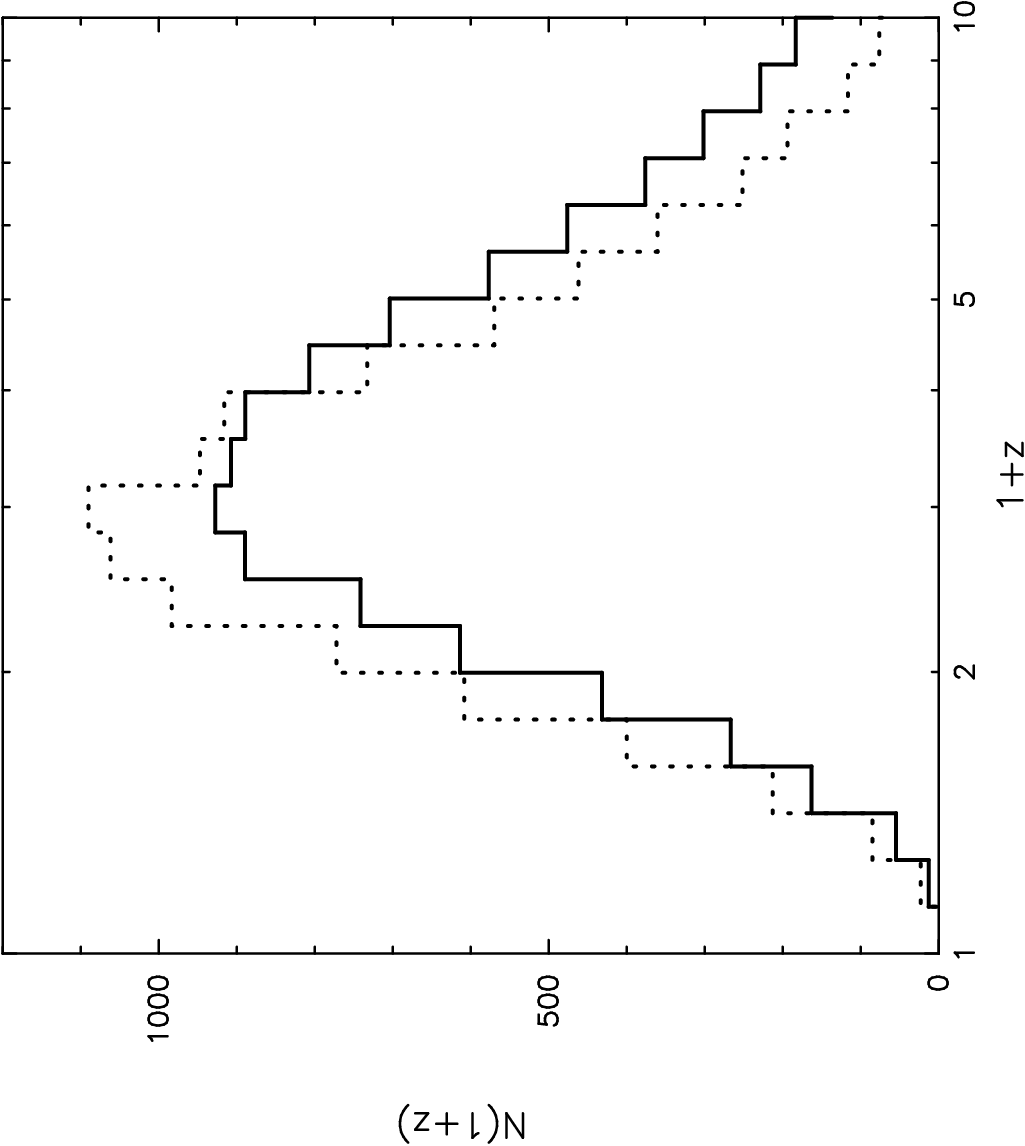}}
\caption{{\it Left:} Comparison of uniform jet models with redshift
evolution (solid) and without (dotted) with the BATSE peak photon number
flux cumulative distribution.  {\it Right:} Predicted observed
distribution of redshifts for the uniform jet model with redshift
evolution (solid) and without (dotted).}
\label{batse}
\end{figure}

\begin{figure}[t]
\begin{minipage}{2.7in}
\includegraphics[width=2.5in,angle=270]{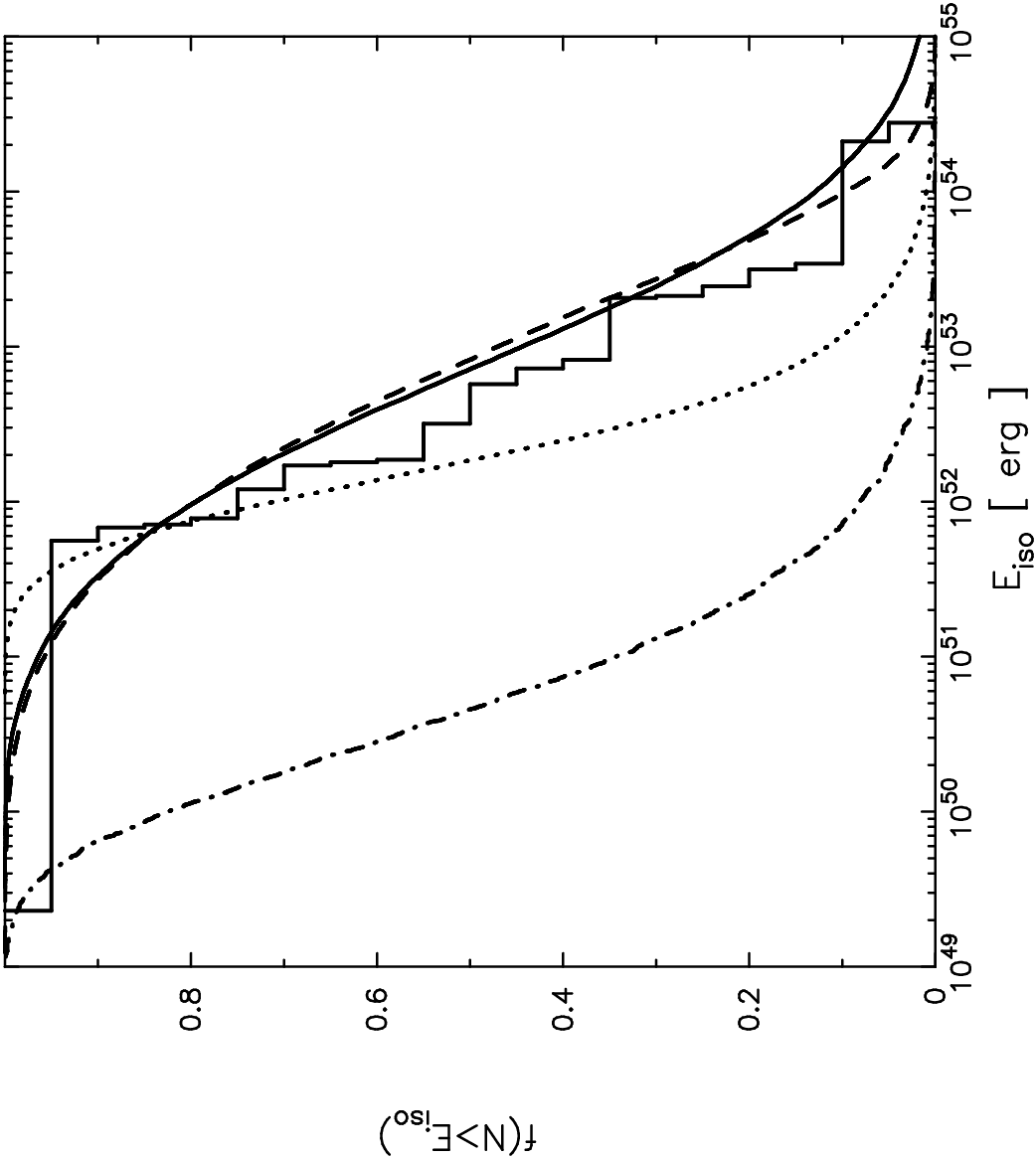}
\\
\includegraphics[width=2.5in,angle=270]{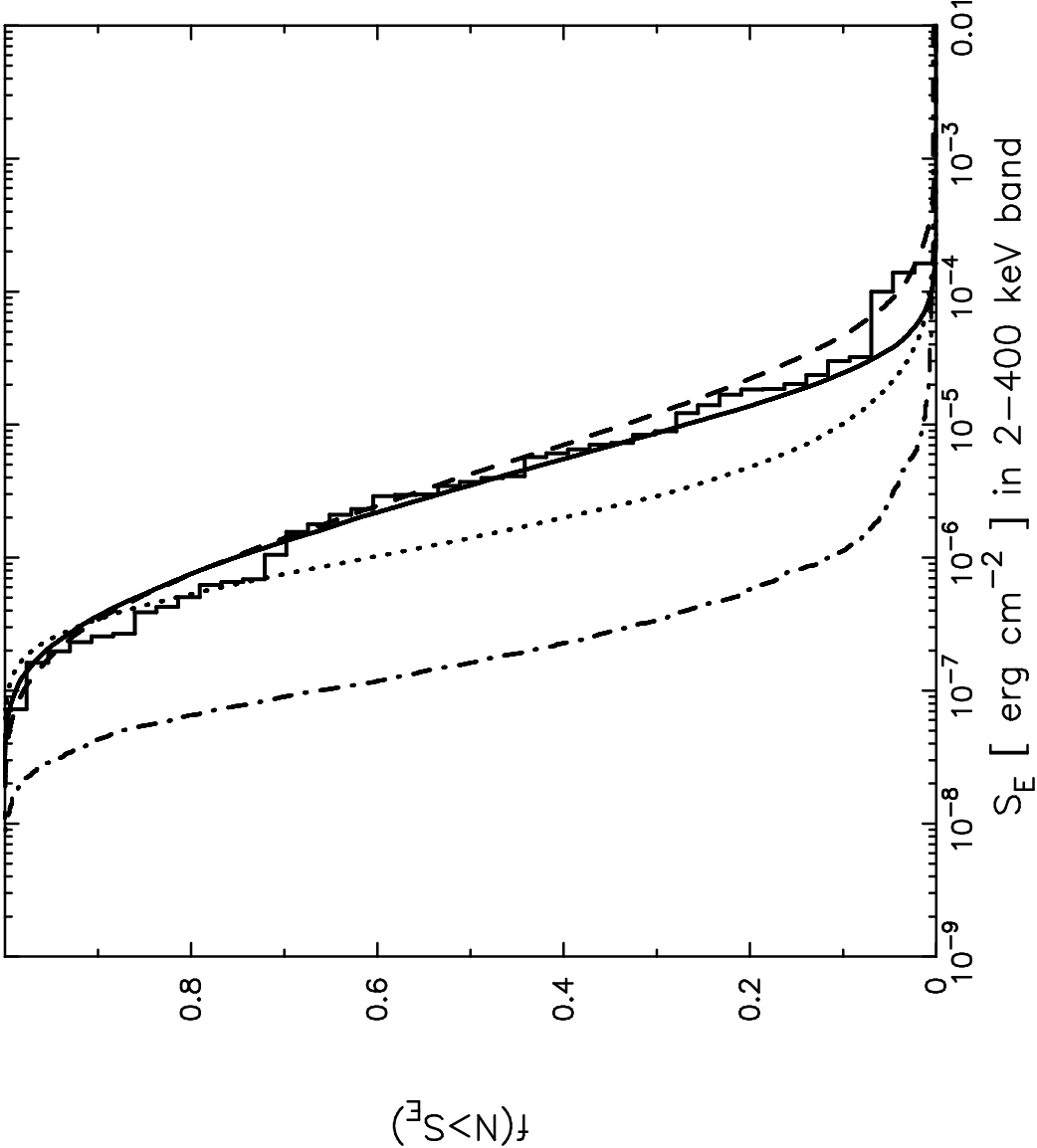}
\end{minipage}
\begin{minipage}{2.7in}
\includegraphics[width=2.5in,angle=270]{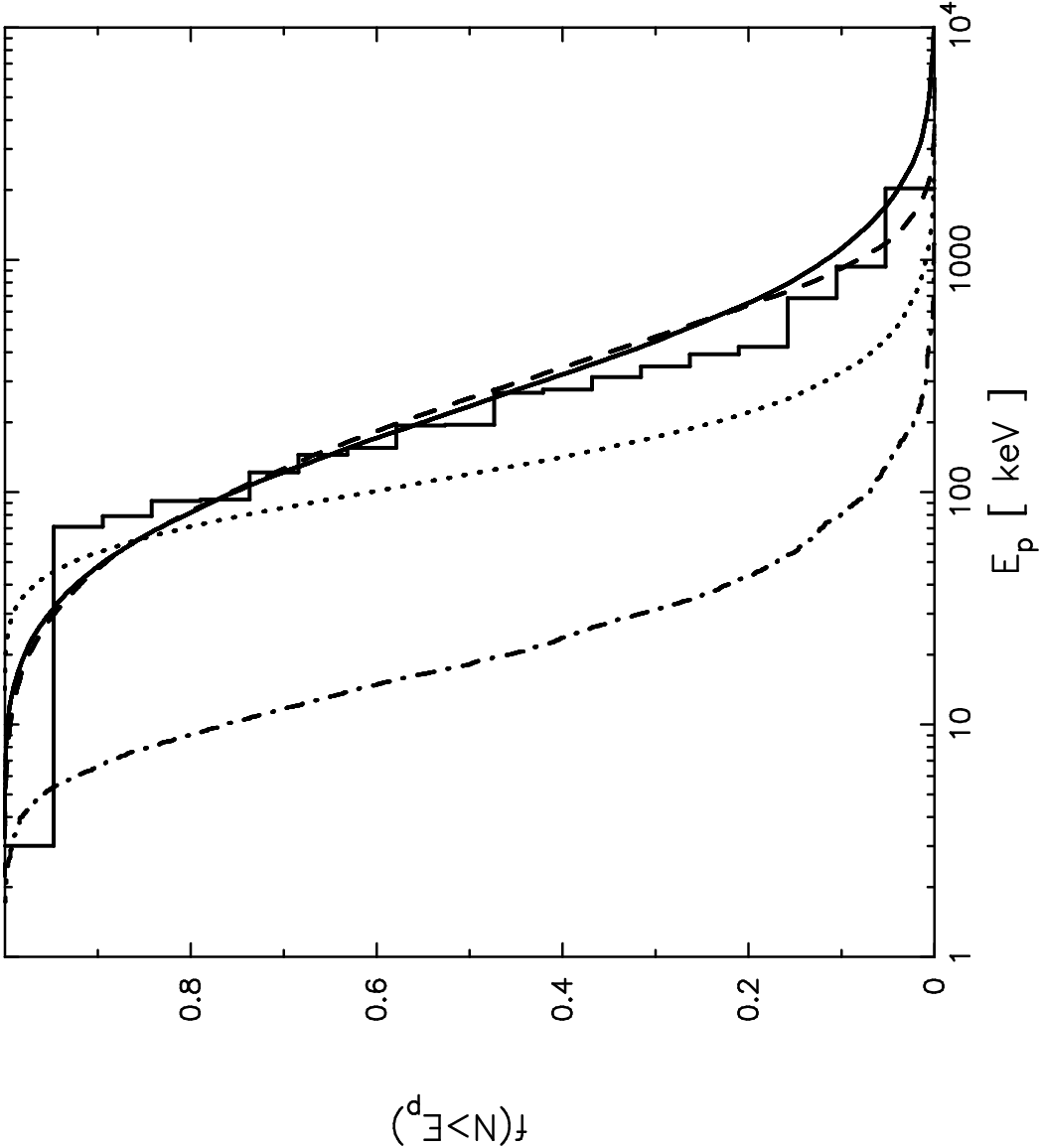}
\\
\includegraphics[width=2.5in,angle=270]{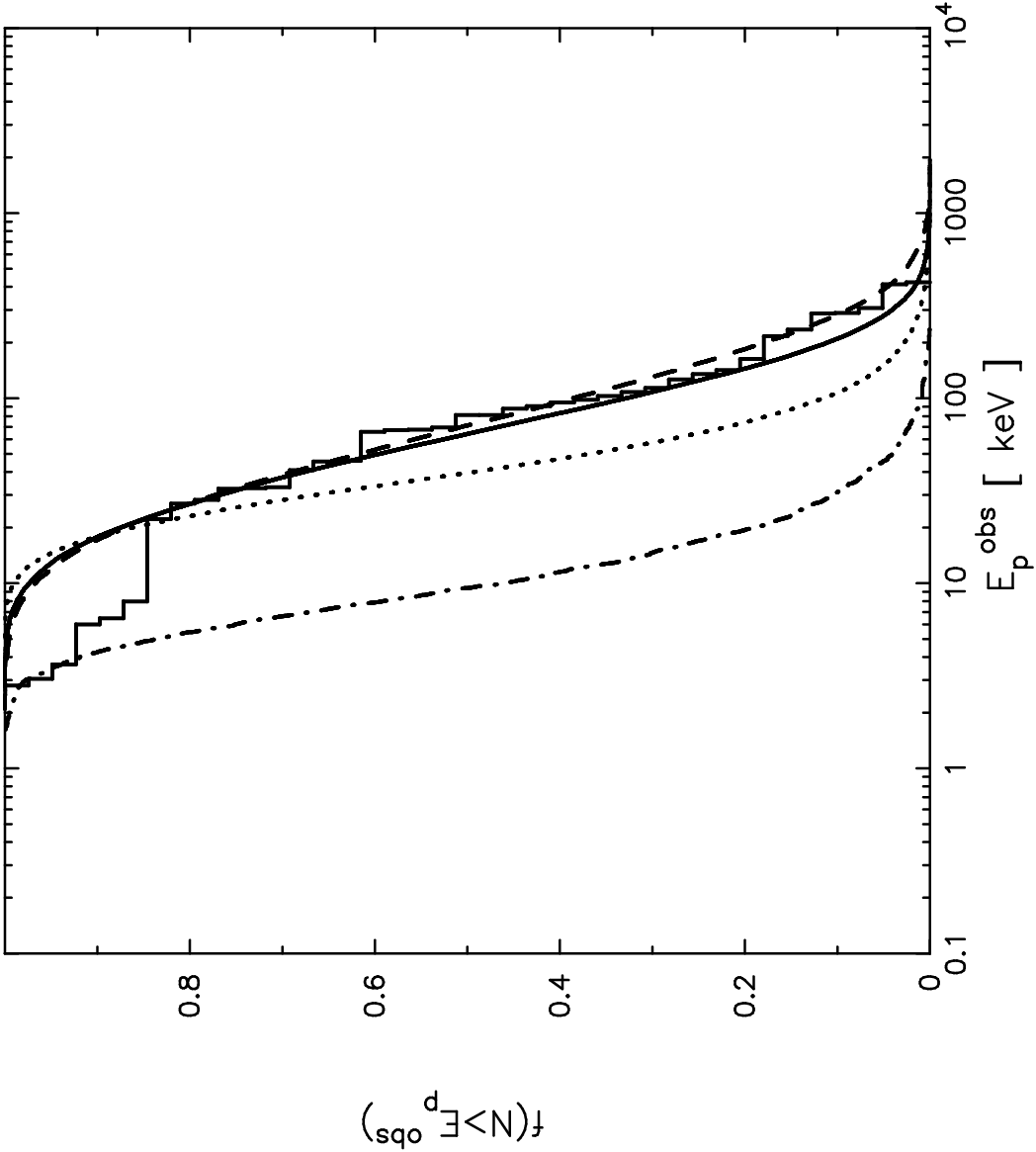}
\end{minipage}
\caption{Comparisons of different models with the observed cumulative
distributions for $E_{\rm iso}$ (upper left), $E_{\rm peak}$ (upper
right), $S_{\rm E}(2-400\; {\rm keV})$ (lower left) and $E_{\rm
peak}^{\rm obs}$ (lower right). The solid curve is the uniform jet
model with redshift evolution and the dashed curve is the uniform jet
model without redshift evolution.  The dotted and dash-dotted curves are
two variants of the universal or structured jet model.}
\label{cuml_dist}
\end{figure}

In addition, adding evolution to the uniform jet model replicates the
observed distribution of peak photon number fluxes as observed by
BATSE.  Models without redshift evolution (Figure \ref{batse}a, dotted
curve) tend to overpredict the number of high peak flux bursts. 
However, models with strong evolution (solid curve) provide excellent
agreement with the BATSE distribution.  Figure \ref{batse}b shows the
differential distribution of redshifts (for bursts detected by WXM) for
the models with and without redshift evolution.

This model makes several predictions.  Most observed XRFs are predicted
to be at $z<1$.  Bursts at low $z$ have $\Omega_{\rm jet} > 10^{-2}$ or
$\theta_{\rm jet} \sim$ a few degrees.  We require a value of the
``standard energy'' to be $E_{\gamma} \sim 5\times 10^{49}$ ergs, or
about $50$ times less than the value reported by \cite{bloom2003}. 
Thus, the fraction of Type Ic supernovae producing GRBs increases from
$\sim 0.1$\% at $z=0$ to $\sim 10$\% at $z \sim 5$.  Finally, $70$\%
of bursts with $z>5$ are detected by the WXM.

\section{Conclusions}

HETE-2 has strengthened the evidence that GRBs evolve with z. The uniform
jet model can describe XRFs and GRBs and can accommodate evolution whereas
the universal jet model cannot.

Confirmation of this model will require the localization of many more
XRFs, the determination of $E_{\rm peak}$ and $E_{\rm iso}$ for many
more XRFs and GRBs, and the identification of optical afterglows and
the measurement of redshifts for these bursts.

HETE-2 is ideally suited to localize XRFs and study their spectra, but
this will be difficult for \swift, which has a nominal threshold of
$E_{\rm min} \sim 15$ keV and a narrow energy band of $15 {\rm \;keV} <
E < 150$ keV.  However, \swift is optimized for pinpointing X-ray and
optical afterglows, and facilitating spectroscopic redshift
measurements.  Therefore, it is very important that the HETE-2 mission
continue, even after \swift is fully operational.  A partnership
between \hetetwo and \swift can confirm or rule out GRB evolution with
redshift.


\begin{thebibliography}{999}
\bibitem{amati2002} Amati, L., et al. 2002, A \& A, 390, 81	

\bibitem{bloom2003} Bloom, J. S., Frail, D. A. \& Kulkarni, S. R. 2003, ApJ, 594, 674

\bibitem{frail2001} Frail, D. A., et al. 200mini1, ApJ, 562, L55

\bibitem{graziani2003} Graziani, C., et al. 2003, in these proceedings

\bibitem{lamb2003} Lamb, D. Q., Donaghy, T. Q. \& Graziani, C. 2003, 
		submitted to ApJ

\bibitem{lamb2003b} Lamb, D. Q., et al. 2003, submitted to ApJ


\bibitem{lloyd2002} Lloyd-Ronning, N. M., Fryer, C. L. \& Ramirez-Ruiz, E.,
2002, ApJ, 574, 565

\bibitem{reichart2001b} Reichart, D. E. \& Lamb, D. Q. 2001

\bibitem{rr2001} Rowan-Robinson, M., 2001, ApJ, 549, 745

\bibitem{sakamoto2003b} Sakamoto, T., et al. 2003, submitted to ApJ

\end{thebibliography}
\end{document}